\documentclass[reprint,aps,prc,twocolumn,superscriptaddress,floatfix,amsmath,amssymb,10pt]{revtex4-2}
\DeclareMathAlphabet{\mathpzc}{OT1}{pzc}{m}{it}
\usepackage{bm}
\usepackage{dcolumn}
\usepackage{amsthm}
\usepackage{amsmath}
\usepackage{amssymb}
\usepackage{graphicx}
\usepackage{braket}
\usepackage{xcolor}
\usepackage{fix-cm}
\usepackage{mathptmx} 
\usepackage[T1]{fontenc}
\usepackage[colorlinks,allcolors=blue]{hyperref}
\makeatletter
\def\NAT@def@citea{\def\@citea{\NAT@separator}}
\makeatother

\begin{document}

\title{Isospin composition of fission barriers}

\author{K. Godbey}\email{godbey@frib.msu.edu}
\affiliation{Facility for Rare Isotope Beams, Michigan State University, East Lansing, Michigan 48824, USA }

\author{Christian Ross}\email{christian.ross@vanderbilt.edu}
\author{A.S. Umar}\email{umar@compsci.cas.vanderbilt.edu}

\affiliation{Department of Physics and Astronomy, Vanderbilt University, Nashville, TN, 37235, USA}

\date{\today}

\begin{abstract}
We employ a microscopic method to study how isospin affect the
fission potential of $^{240}$Pu. Our approach uses constrained
Hartree-Fock theory (CHF) which allows us to separately investigate the
isoscalar and isovector properties of the nuclear energy density functional (EDF).
By analyzing the isoscalar and
isovector components of the EDF along the fission path
we can assess the isovector contribution to fission barriers. We study this effect for the fully adiabatic path to scission.
The isovector component of the fission potential is found to increase in magnitude as
the nucleus evolves towards scission, exemplifying the importance of stringent
constraints on the isovector sector of the nuclear EDF for reliable predictions
of fission properties.

\end{abstract}

\maketitle

{\it Introduction.}
Fission occurs in various complex quantum systems, such as atomic nuclei and atomic clusters. A robust theoretical formulation of the fission process represents a significant challenge in the field of quantum many-body physics, primarily because of the difficulties in developing an effective and computationally feasible approach to model the transition from a unified system to the creation of separate fragments~\cite{schunck2016,andreyev2017,bender2020,schunck2022}.
In the context of atomic nuclei, most of the evolution is believed to unfold over a relatively long time scale.
This comparatively slow progression allows for the initial treatment of nuclear fission as an adiabatic process, where the  system transitions along a potential energy landscape characterized by macroscopic variables like shape elongation and asymmetry.

Microscopic methods have been extensively applied in the analysis of fission pathways, as evidenced by recent studies in references\cite{dubray2008,goutte2005,bonneau2006,staszczak2009,pei2009a,younes2011,warda2012,abusara2012,mirea2012,lu2012,staszczak2013,mcdonnell2013,sadhukhan2013,schunck2014,verriere2020,bernard2021,bernard2023,flynn2022,lay2024,qiu2024}.
Most microscopic approaches to calculate static adiabatic fission potentials rely on mean-field approaches with the effective interaction given by a nuclear energy density functional (EDF).
The majority of EDFs are determined for ground state properties of nuclei with neutron-proton asymmetries ranging
from small to moderate values, leaving the isovector parts of the EDFs relatively under-constrained~\cite{godbey2022}.
The isovector part of the EDF
plays an important role in determining properties of atomic nuclei, particularly those
with large neutron-proton asymmetry~\cite{bender2003,vretenar2005,stone2007,sheikh2014}.
In addition to influencing nuclear structure properties such as
isovector giant dipole resonances and beta decay rates
the isovector part also significantly effects the
neutron skin thickness for heavy nuclei, which is important in
constraining the symmetry energy and consequently impacting neutron stars and
other astrophysical phenomena such as nucleosynthesis~\cite{li2008,baldo2016}.
Particularly for neutron-rich nuclei, the isovector characteristics of the interaction will become increasingly significant.
Recently, we have shown that the isovector part of the EDF plays an important role in the microscopic
calculation of fusion barriers~\cite{godbey2017}. In the case of fusion, the isovector contribution is critical to the $N/Z$ equilibration
that occurs when the two nuclei begin to interact~\cite{simenel2020}, which in turn modifies the fusion barrier.
Here, we adopt the same formalism to study isovector contribution of the EDF to the adiabatic fission potential of $^{240}$Pu.
Next we discuss the methods used to obtain the fission barriers as well as the isoscalar and isovector decomposition. This is followed by the discussion of our results.

{\it Formalism.}
\label{formalism}
Within the Hartree-Fock (HF) theory, the antisymmetric many-body wavefunction is taken to be a
Slater determinant. This many-body state is
then used to construct the action using an effective nucleon-nucleon
interaction. Variation of this action with respect to single-particle states
\begin{equation}
    \label{eq:variation}
    \delta S = \delta\int \braket{\Phi | \hat{H} - \lambda (\hat{Q}-Q_0) |\Phi} = 0,
\end{equation}
subject to a constraint operator $\hat{Q}$ with the Lagrange multiplier $\lambda$, results in the constrained Hartree-Fock (CHF) equations, which forces the system to acquire a particular value of $Q_0=\braket{\Phi_0 |\hat{Q}|\Phi_0}$,
where the states $\Phi_0$ are the solutions of Eq.~(\ref{eq:variation}) for a particular choice of $Q_0$.
The numerical solution of the CHF equations using the gradient iteration method is given in Refs.~\cite{umar1985,cusson1985,umar2012}.
If we employ an effective interaction such as the
Skyrme interaction, the total energy of the system can be represented as an
volume integral of an EDF~\cite{engel1975}
\begin{equation}
    \label{eq:energy}
    E =  \braket{\Phi | \hat{H} |\Phi} = \int d^3\mathbf{r} {\cal{H}}(\mathbf{r})~,
\end{equation}
with $\braket{\Phi | \Phi}=1$.
The Skyrme EDF may be decomposed into isoscalar and isovector parts~\cite{dobaczewski1995}
(in addition to the conventional kinetic and Coulomb terms) as:
\begin{equation}
    \label{eq:edensity}
    {\cal{H}}(\mathbf{r}) = \frac{\hbar^2}{2m}\tau_0
    + {\cal H}_0(\mathbf{r})
    + {\cal H}_1(\mathbf{r})
    + {\cal H}_C(\mathbf{r})~.
\end{equation}
The isoscalar and isovector terms carry an isospin index ($I = 0, 1$) for the energy densities, respectively.
The isoscalar energy density (${\cal H}_0(\mathbf{r})$) depends on the isoscalar particle density,
whereas the isovector energy density (${\cal H}_1(\mathbf{r})$) depends on the isovector particle
density, which can be defined through the density operator
\begin{equation}
\hat{\rho}_{\mathrm{I}}(\mathbf{r}) = \sum_{\lambda=1}^{A} \delta(\mathbf{r}-\mathbf{r}_{\lambda})\hat{a}_{\mathrm{I}}(\lambda)\;,
\end{equation}
where
\begin{equation}
<q|\hat{a}_\mathrm{I}|q'>=\left\{ \begin{aligned}  (-1)^{\mathrm{I}}&\delta_{qq'},\; q=\mathrm{proton} \\
                 &\delta_{qq'},\; q=\mathrm{neutron} \end{aligned} \right. \;,
\end{equation}
allowing us to define
\begin{equation}
    \rho_{\mathrm{I}}(\mathbf{r})=\left\{ \begin{aligned} &\rho_n(\mathbf{r})+ \rho_p(\mathbf{r}),\;\; \mathrm{I}=0 \\
                 &\rho_n(\mathbf{r})- \rho_p(\mathbf{r}) ,\;\; \mathrm{I}=1\end{aligned} \right. \;.
\
\end{equation}
These definitions, of course, prescribe analogous expressions for other densities and
currents. Using the above definition various moments are given by
\begin{equation}
    (q_{L0})_{\mathrm{I}} = \int\, d^3r\, r^L\,Y_{L0}(\hat{\mathbf{r}})\,\rho_{\mathrm{I}}(\mathbf{r})\;.
\end{equation}

The local gauge and Galilean invariant form of the EDF is given by~\cite{dobaczewski1995}
\begin{equation}
    \begin{aligned}
        \label{eq:efunctional}
        {\cal H}_\mathrm{I}(\mathbf{r})
        & = C_\mathrm{I}^{\rho}            \rho_\mathrm{I}^2
        +  C_\mathrm{I}^{   s}            \mathbf{s}_\mathrm{I}^2
        +  C_\mathrm{I}^{\Delta\rho}      \rho_\mathrm{I}\Delta\rho_\mathrm{I} \\
        &+  C_\mathrm{I}^{\Delta s}        \mathbf{s}_\mathrm{I}\cdot\Delta\mathbf{s}_\mathrm{I}
        +  C_\mathrm{I}^{\tau}      (\rho_\mathrm{I}\tau_\mathrm{I}-\mathbf{j}_\mathrm{I}^2) \\
        &+  C_\mathrm{I}^{   T}      \Big(\mathbf{s}_\mathrm{I}\cdot
        \mathbf{T}_\mathrm{I} - \tensor{J}_\mathrm{I}^2\Big)
        + C_\mathrm{I}^{\nabla J}  \Big(\rho_\mathrm{I}\mathbf{\nabla}\cdot\mathbf{J}_\mathrm{I}
        + \mathbf{s}_\mathrm{I}\cdot
        (\mathbf{\nabla}\times\mathbf{j}_\mathrm{I})\Big)\;.
    \end{aligned}
\end{equation}
The density dependence of the coupling constants has been restricted to the $C_\mathrm{I}^{\rho}$ and $C_\mathrm{I}^s$ terms only
which stems from the most common choice of Skyrme EDF.
These density dependent coefficients contribute to the coupling of isoscalar and isovector fields
in the Hartree-Fock Hamiltonian~\cite{dobaczewski1995}.
In static CHF calculations for even-even nuclei the time-odd terms, $\mathbf{s}$ and $\mathbf{j}$
in Eq.~(\ref{eq:efunctional}), are zero due to time-reversal invariance. Similarly, most Skyrme parametrizations
do not include the $C_\mathrm{I}^T$ term in the functional.

The decomposition of the Skyrme EDF into isoscalar and isovector components makes it feasible to study isospin dependence of nuclear properties microscopically, both for nuclear reactions~\cite{vophuoc2016,godbey2017,gumbel2023} as well as for nuclear structure~\cite{dobaczewski1995}.
This is possible for any approach that employs the Skyrme EDF. Here,
we implement the decomposed Skyrme EDF to study
isospin effects in fission barriers coming from a constrained approach.
Utilizing the decomposition of the Skyrme EDF [Eqs.~(\ref{eq:edensity} and \ref{eq:efunctional})], we can re-write the energy along the
constraint path as
\begin{equation}
    E(Q_0) = \sum_{\mathrm{I}=0,1} E_\mathrm{I}(Q_0)~,
\end{equation}
where $E_\mathrm{I}(Q_0)$ denotes the energy of the system computed by using the isoscalar and isovector parts of
the Skyrme EDF given in Eqs.~(\ref{eq:edensity}) and (\ref{eq:efunctional}). The isoscalar energy term also includes the
contributions from the kinetic energy and Coulomb terms.
The Coulomb potential is solved from the typical three-dimensional Poisson equation (where the Slater approximation is used for the Coulomb exchange term) via Fast-Fourier Transform techniques.
In practice it is more convenient to subtract the ground state energies to define the fission potential
as
\begin{eqnarray}
    V_I(Q_0) &=& E_\mathrm{I}(Q_0)-E_\mathrm{I}^{g.s.}(Q_0) \nonumber \\
    V(Q_0)   &=& \sum_{\mathrm{I}=0,1} V_{\mathrm{I}}(Q_0)~.
\end{eqnarray}

{\it Results.}
\label{results}
Applying the formalism detailed above, we now present the adiabatic fission potentials of $^{240}$Pu using the CHF
method.
This nucleus is well-motivated as it has been shown that data on its fission isomer strongly impacts the resulting EDF calibration, particularly in the symmetry energy and its slope~\cite{kortelainen2012}.
The Skyrme energy density functional SLy4d~\cite{kim1997} was used with pairing treated at the BCS level.
The pairing scheme was adopted to be the same as in the EV8 code~\cite{ryssens2015}, a density dependent surface pairing contact interaction
with Fermi function cutoffs~\cite{bender2000,ryssens2015,dobaczewski1995b}. This was done
to allow for comparison
of the results calculated with the EV8 code.

The code used here was the VU-TDHF code which assumes
no quantal or geometrical symmetries or approximations and includes the full Skyrme interaction,
including the time-odd terms~\cite{umar1991a,umar2006c}. Unrestricted symmetry calculations of the
fission potential is much more computationally intensive in comparison to axially symmetric
or three-dimensional codes that impose planar symmetries.
The VU-TDHF code uses basis-spline discretization and a gradient
iteration method~\cite{bottcher1989} for highly accurate calculations at a reasonable computational cost. The numerical implementation of the constrained
iterations are described in Refs.~\cite{umar1985,umar2012}. While we do have possibility of imposing multiple constraints,
we have only used the quadrupole constraint
 $$q_{20}=\sqrt{\frac{5}{16\pi}}\int d^3r \,\rho(\mathbf{r}) (2z^2-x^2-y^2),$$
in steps of 100~$fm^2$ starting from the ground state, with
deformation 907~$fm^2$, up to 12000~$fm^2$ (note that some studies do not include the factor
$\sqrt{5/16\pi}$ in the definition of the quadrupole constraint).
This allows the system to find the minimum energy configuration given the defined elongation of the nucleus.
These configurations may involve arbitrary deformations, including triaxiality as discussed later.
At the final constraint step the fragments are separated by
$R=17.7$~fm.
\begin{figure}[!hbt]
    \includegraphics[width=8.6cm]{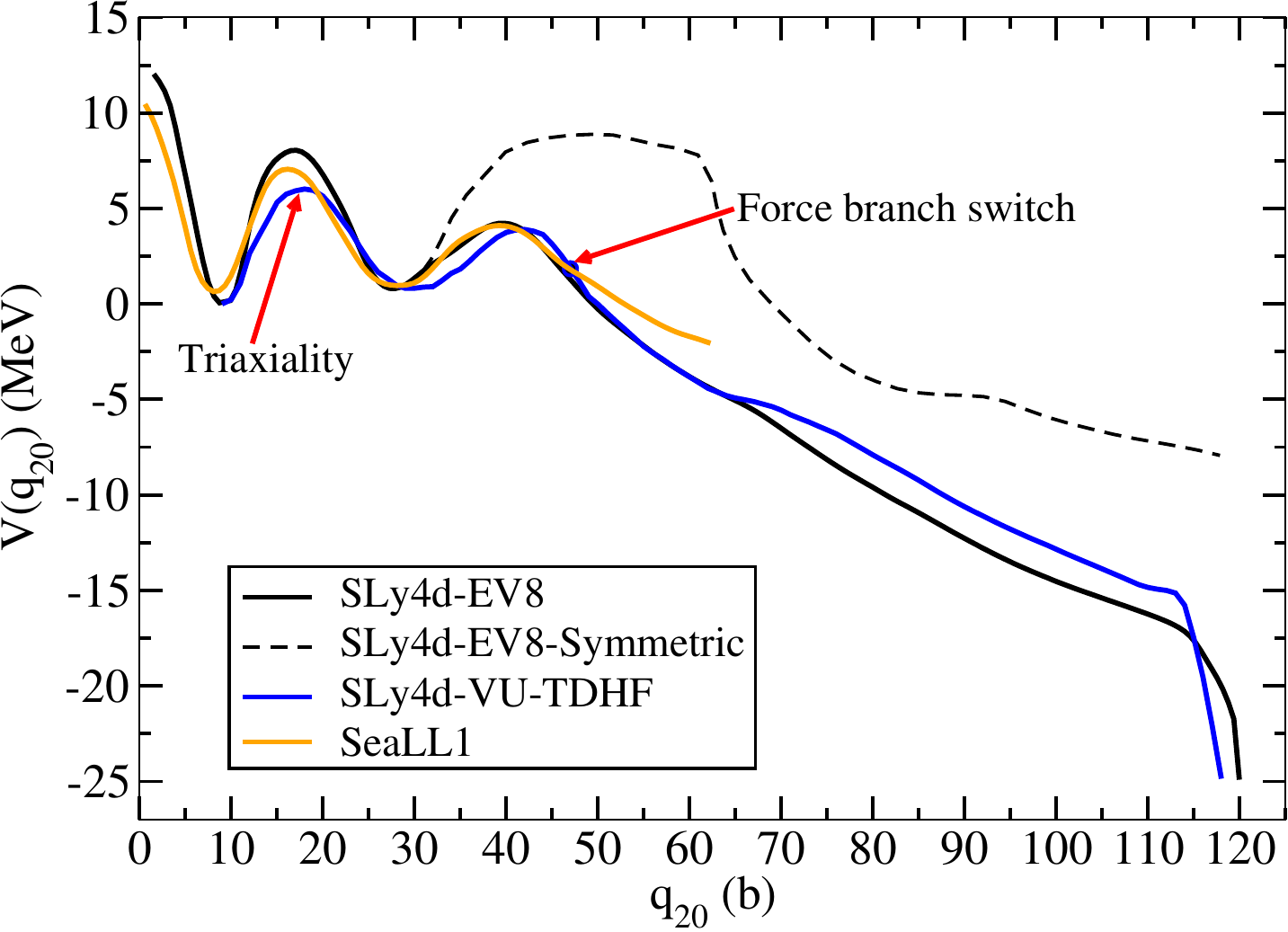}
    \caption{\protect Fission potentials for $^{240}$Pu calculated with different codes and functionals.
    Our results are depicted by the blue curve, while the EV8 results for asymmetric fission is shown with
    the solid black curve~\cite{priv_scamps}. The symmetric fission potential of EV8 is shown by the dashed black curve.
    For comparison we have also plotted the fission potential using the
    SeaLL1 EDF of Ref.~\cite{bulgac2018} as the orange color curve.
    }
    \label{fiss}
\end{figure}
The final light fragment average mass and charge is found to be $A=107.71$ and $Z=43.28$ with corresponding
values for the heavy fragment being $A=132.29$ and $Z=50.72$. These values are in agreement with
those found in a recent study using the constrained Hartree-Fock Bogoliubov (CHFB) method~\cite{tong2022b}.
The numerical box size was $50\times32\times32$~fm with mesh spacing of 1~fm.
As an additional check we have also performed a backward constraint, starting from the final separated fragments,
which coincided with the same curve.

In Fig.~\ref{fiss} we show the fission potentials for $^{240}$Pu calculated with different codes and in one case a different EDF.
Our results are depicted by the blue curve, while the fission potential calculated for the asymmetric fission using the EV2 code, a more general version of EV8~\cite{ryssens2015} that includes the possibility of parity breaking, is shown with the
solid black curve~\cite{priv_scamps}. The symmetric fission potential of EV8 is shown by the dashed black curve.
For comparison we have also plotted the fission potential using the
SeaLL1 EDF of Ref.~\cite{bulgac2018} as the orange color curve.
The fully unrestricted constraint procedure starts with an axially symmetric ground state with no octupole deformation
and continues as such until about 4700~$fm^2$. Beyond this point we move towards the symmetric fission channel. However,
by tightening the constraint convergence criteria, and after many iterations, the system falls back onto the asymmetric branch developing an octupole deformation.
The lower first barrier height is due to the presence of triaxiality~\cite{abusara2010,sadhukhan2014,benrabia2017} that is explored in our calculations as compared to the results from the EV8 code, where the calculations where limited to axial
symmetry for ease of use.
We also observe that the switch to the symmetric fission branch occurs later in our case for the same reason.
\begin{figure}[!hbt]
    \includegraphics[width=8.6cm]{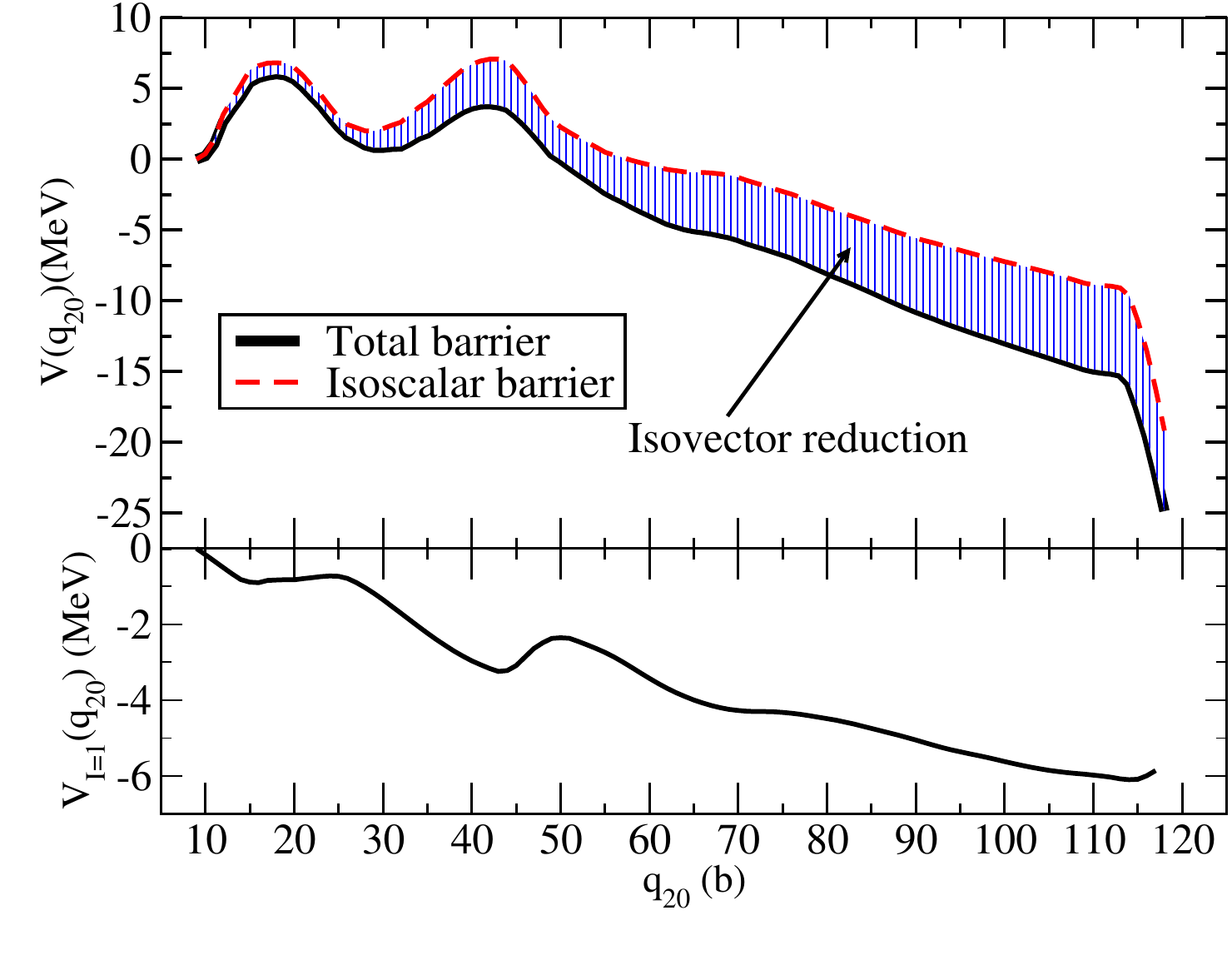}
    \caption{\protect Total (solid line) and isoscalar (dashed line) fission barriers for $^{240}$Pu is shown on the upper panel. The shaded
    region depicts the reduction of the fission barrier due to isovector contribution. The bottom panel
    shows the isovector potential as well as the potential corresponding to time-dependent scission case
    (dotted line).}
    \label{fiss-iso}
\end{figure}

In the upper panel of Fig.~\ref{fiss-iso} we plot the total fission barrier as well as the one originating
specifically from the isoscalar part of the EDF. The shaded
region shows the reduction of the total barrier due to isovector contribution.
This reduction of the potential leads naturally to an enhancement in overall fission probabilities.
The bottom panel of Fig.~\ref{fiss-iso} shows the isovector contribution only. We observe that the isovector
contribution acts to lower the isoscalar barrier throughout the fission potential curve, resulting in a
lower total fission potential.
\begin{figure*}[!htb]
    \includegraphics*[width=14cm]{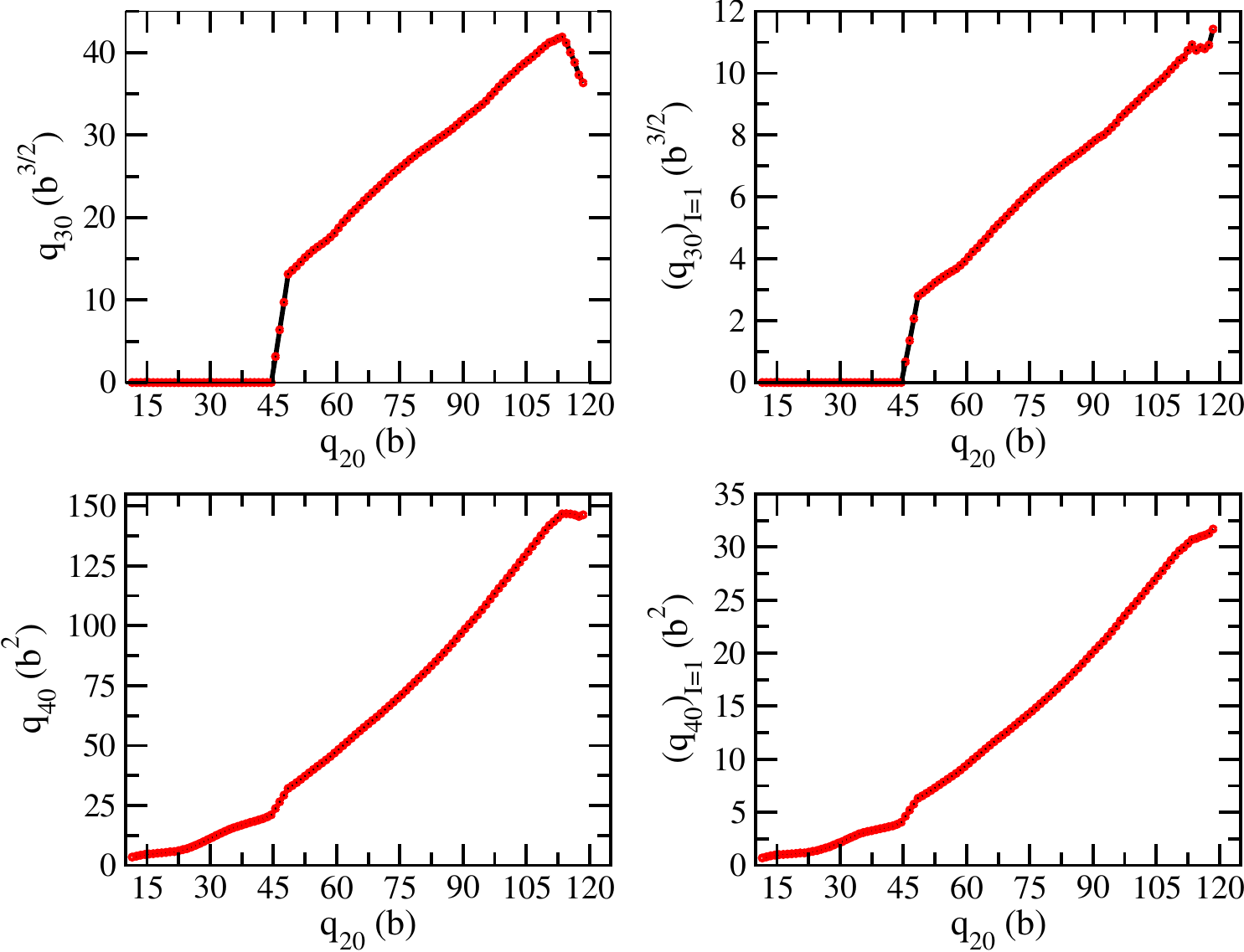}
    \caption{\protect The evolution of total and isovector octupole and hexadecupole moments
    as a function of the total quadrupole moment.}
    \label{fiss-qxx}
\end{figure*}

In Fig.~\ref{fiss-qxx} we plot the variation of the total and isovector octupole and hexadecupole moments as a function of the total quadrupole moment. As we see from the top panels the system
starts with an axially symmetric configuration with no octupole deformation and in the vicinity
of $q_{20}=4500$~fm$^2$ it makes a relatively rapid transition to acquire an octupole deformation.
This is the case for both total and isovector octupole moments with isovector part having a
lower strength. The transition to acquire an octupole deformation is somewhat dependent on the
theoretical approach used to calculate the fission potential, with full HFB calculations
yielding a smoother transition~\cite{tong2022b}.
The hexadecupole moment has a smoother evolution through the transition point
and again grows in a somewhat linear fashion with increasing quadrupole moment.

Over the years various experiments have been done to measure the height and depth of fission barriers for the
$^{240}$Pu nucleus~\cite{specht1974,vandenbosch1977,metag1980,bjornholm1980,thirolf2002,singh2002}.
Bases on these the inner fission barrier height is predicted to be approximately 6.05~MeV,
second minimum around 2.8~MeV, and outer fission barrier
height to be around 5.15~MeV. In our calculations the isoscalar inner barrier height is 6.6~MeV, the first minimum
depth is 2.2~MeV, and the outer barrier height is 6.0~MeV.
The corresponding total potential values are; inner barrier height
of 6.0~MeV, first minimum depth of 0.82~MeV, and outer barrier height of 3.9~MeV.
From these values we see that the isovector contribution significantly reduces the the depth of the
inner barrier to be much lower than the experimental value as well as a significant reduction in
the height of the outer barrier. The very low inner barrier depth is a common problem in most
studies using various EDFs.

{\it Discussion.}
\label{discussion}
There is growing evidence that the isovector part of the nuclear energy
density functionals play an important role for a multitude of structure
and reaction properties. With the increasing availability of neutron and proton rich
nuclei produced in radioactive ion-beam facilities more data should
be available to better ascertain the contribution of the isovector
components to the EDF. The isovector contributions do not only arise
from the core EDF but also from the spin-orbit~\cite{sharma1995,yue2024} and pairing parts~\cite{yamagami2012}.
We have undertaken a study to discern the isoscalar and isovector contributions to the one-dimensional fission
potential.
The fission potential was calculated using a three-dimensional solver with no
symmetry restrictions and
the SLy4d EDF was used to facilitate comparison with other calculations.
A natural future extension to this approach would be to extend the analysis to studies of potential energy surfaces and systems that exhibit multimodal fission, though this represents a substantial increase in computational effort.

We observe that the isovector part of the EDF makes a substantial contribution to the fission
potential along the fission trajectory, accumulating influence as the system nears the scission point.
This raises the question as to how well we constrain the isovector part of the
EDF, including the isovector dependence of the spin-orbit interaction.
That fusion and fission properties are important to constraining this sector is not new, though the consistent increase along the fission pathway implies that observables that are strongly impacted by the elongation of the system at scission are likely to contain more information than the first barrier and fission isomer alone.
Naturally, any improvement
to the isovector EDF would necessitate simultaneous consideration of the isoscalar part as well.
Finally, a proper treatment of superfluidity, especially for the scission process that benefits from
a dynamical pairing approach such as the time-dependent HFB, may introduce further improvements
to pin down the isovector properties of the EDF.
During the dynamical scission and final phase of fragment formation, the rearrangement of nucleons due to shell effects influencing
the $N/Z$ equilibration will become very important in a way somewhat similar to the initial stages of the fusion process.
Furthermore, some of the terms that vanish in the static calculations become relevant during the dynamics, meaning that observables sensitive to those dynamics (such as the total kinetic energy) are the only hope one has of providing a strong constraint.

\begin{acknowledgments}
This work has been supported by the U.S. Department of Energy under award Nos. DE-SC0013847 (Vanderbilt University) and DE-NA0004074 (NNSA, the Stewardship Science Academic Alliances program).
\end{acknowledgments}

\bibliography{VU_bibtex_master}

\end{document}